%% file: main.tex
\documentclass[%
 reprint,
 amsmath,amssymb,
 aps,
]{revtex4-1}

\usepackage{graphicx}% Include figure files
\usepackage{dcolumn}% Align table columns on decimal point
\usepackage{bm}% bold math
\usepackage{framed}

\newcommand{\xcal}{\cal X}

\newcommand{\expX}[1]{\langle #1 \rangle}

\newcommand{\varX}{\text{ Var}\, X}

\begin{document}

\title{Measuring complexity}% Force line breaks with \\
%\thanks{A footnote to the article title}%

\author{Karoline Wiesner}
 \affiliation{School of Mathematics, University of Bristol, U.K.}
 \email{k.wiesner@bristol.ac.uk}

\author{James Ladyman}
\affiliation{
 Department of Philosophy, University of Bristol, U.K.
}%

\date{\today}% It is always \today, today,
             %  but any date may be explicitly specified

\begin{abstract}
Complexity is a multi-faceted phenomenon, involving a variety of features including disorder, nonlinearity, and  self-organisation. We use a recently developed rigorous framework for complexity to understand measures of complexity. We illustrate, by example, how features of complexity can be quantified, and we analyse a selection of purported measures of complexity that have found wide application and explain whether and how they measure complexity. We also discuss some of the classic information-theoretic measures from the 1980s and 1990s. This work gives the reader a tool kit for quantifying features of complexity across the sciences.
\end{abstract}

%\keywords{Suggested keywords}%Use showkeys class option if keyword
                              %display desired
\maketitle

%\tableofcontents

\include{body}

\include{appendix}

\end{document}

%% file: body.tex
\begin{framed}
    \textit{Sections II -- XI of this text are, with small modifications, a chapter of our book ``What is a complex system?'', published with Yale University Press \cite{ladyman2020what}. We decided to make this chapter available separately as it requires mathematical knowledge, in contrast to the rest of the book. Please always cite this article together with the book.} 
    \end{framed}

Many measures of complexity have been proposed since scientists first began to study complex systems, and the list is still growing. 
Complexity is a multi-faceted phenomenon, and complex systems have a variety of features not all of which are found in all of them. This implies that assigning a single number to complexity cannot do it justice. As the late physics Nobel laureate Murray Gell-Mann noted early on, ``A variety of different measures would be required to capture all our intuitive ideas about what is meant by complexity'' \cite[p.1]{gell-mann_what_1995}. This article is a guide to quantifying complexity while respecting this insight.

The fact that any measure measures only an aspect of complexity and never the phenomenon as a whole should inform any practitioner's approach to quantifying complexity to avoid unnecessary disagreement and bring clarity to the basic claims and concepts of the field, and to allow measures of different kinds can be used together.

We begin by summarising the features of complex systems identified in \cite{ladyman2020what}. In the main part of this article, we discuss mathematical means available to quantify each of them them accompanied by examples. A feature can take different forms in different complex systems and scientific domains (and even within a single complex system) and, therefore, there is often more than one way by which to measure any given feature. The techniques were often invented in a different discipline from those in which they are now applied as the provided examples from the scientific literature illustrate.

Since complexity is a collection of features rather than a single phenomenon, all quantitative measures of complexity can quantify only aspects of complexity rather than complexity as such. Accepting this insight makes it necessary to ask what any purported `measure of complexity' actually measures. In the final section of this article, a few, by now classic, measures of complexity from the 1980s and 1990s, mentioned in many discussions on the subject, are discussed, including effective complexity, effective measure complexity, statistical complexity, and logical depth. 
These `classic' measures were constructed as thought experiments rather than as measures to be applied to real-world systems. Hence, they are not tools that can be used in practice. However, they have played an important role in the development of our understanding of complexity over the decades. They now help us understand the distinction between measuring complexity and measuring features of complexity. 

Throughout, terminology is explained when it is used for the first time. Further mathematical background is given in the Appendix. 

\section{What is a complex system?}
\label{sec.features}

In recent work, we have developed a framework for understanding `complexity' which is applicable across the natural and social sciences \cite{ladyman2020what}. We have distilled a list of features that are exhibited by complex systems; some features are exhibited by all complex systems, some only by functional or living complex systems. We distinguish between conditions for complexity and products of complexity. In a nutshell, the products are the `emergent' properties that arise because of the many disordered interactions between the many parts and the feedback from previous interactions in systems that are open to the environment in some way. The latter are the `conditions for complexity' (numerosity of elements and interactions, disorder, feedback, non-equilibrium). Table~\ref{tab.features} lists the features of complexity identified. Not all products are present in all complex systems. In particular, as mentioned above, some are only present in functional or living complex systems (robustness of function, adaptive behaviour, modularity, memory). This is true by definition of these properties. Examples of non-functional / non-living systems are the universe and many condensed-matter  systems, in particular when they exhibit phase transitions.

There is an important distinction between the \emph{order of} a complex systems and the \emph{order produced by} a complex system. An example of order produced by a complex system is a snowflake  produced by the weather and climate system. Complex systems are always dynamic, but they often produce static order. Another example of a complex system that produces order is a honey bee hive; the \emph{order of} the hive is the self-organised patterns of labour distribution for example; the (static) order \emph{produced by} the hive are honey combs with their intricate hexagonal structure. 
In short, a complex system is a system that exhibits all of the conditions for complexity and at least one of the products emerging from the conditions.  
Here, we will not discuss these features much further. For details, see \cite{ladyman2020what}. 

\begin{table}
  \caption{\label{tab.features} The features of complexity, as identified in \cite{ladyman2020what} where they have been grouped into `conditions for complexity' and `products of complexity'.}
  \begin{center}
  \begin{tabular}{ | l |}
	\hline 
	\textbf{Conditions for complexity} \\
	\hline
	Numerosity of elements \\
	Numerosity of interactions \\
	Disorder \\
	Non-equilibrium (openness) \\
	Feedback \\
	\hline 
	\hline 
	\textbf{Products of complexity} \\
	\hline
	Nonlinearity\\
	Self-organisation\\
	Robustness of order\\
	Nestedness\\
	Robustness of function\\
	Adaptive behaviour\\
	Modularity\\
	Memory \\
	\hline 
  \end{tabular}
  \end{center}
\end{table}

We now go through each feature in Tab.~\ref{tab.features} and give examples of measures that quantify them.

\section{Numerosity} 

The most basic measure of complexity science is the counting of entities and of interactions between them. Numerosity is the oldest quantity in the history of science, and counting is among the most basic scientific methods. Counting is the foundation of measurement because quantities of everyday relevance such as length, mass and time can be counted in units such as metres, grams and seconds. Counting alone does not tell us what counts as `more' in the sense of `more is different', because, when we consider emergent behaviour, how many is `enough' depends on the system. 
For some systems it is the high number of elements that is relevant for complexity, as in fluid dynamical systems; for others it is the high number of interactions, as in a small group of swarming animals or small insect colonies; or it is both, as in the brain and many (if not most) complex systems. The number of interactions is as important as the number of elements in the system. 

\section{Disorder and Diversity}
\label{sec.disorder.measure}

Disorder and diversity are related, and the words used to describe them overlap and are often not clearly defined. `Disorder' usually refers to randomness, which is to say lack of correlation or structure. Disorder is therefore just the lack of order. It is worth stating this explicitly since it follows that any measure of order can be turned into a measure of disorder and vice versa. 

A disordered system is one that is random in the sense of lacking correlations between parts over space or time or both, at least to some extent. It is worth remembering that complex systems are never completely disordered. In complex systems, disorder can exist at the lower level in terms of the stochasticity in the interactions between the parts, as well as at the higher level, in terms of the structure which emerges from them and which is never perfect. Thermal fluctuations are a form of disorder relevant to the dynamics of complex systems. For example, thermal fluctuations are necessary for most biochemical processes to take place. The term `noise' or `thermal noise' is used more frequently than `disorder' in this context.

A real or purely mathematical random system would not be described as `diverse'. Instead, the term `diversity' is often used to describe inhomogeneity in element type -- that is types of different kinds. Measures have been designed specifically to  address diversity in this sense. Some of these are discussed at the end of this section. 

Interactions can be disordered in time or in their nature. Elements can be disordered in terms of type. The structure formed by a complex system can be disordered in its spatial configuration. All these kinds of disorder are relevant, and all are quantifiable. 

Mathematically, disorder is described with the language of probability (see  Appendix~\ref{sec.probtheory} for a brief introduction to probability theory). The elements or interactions which are disordered are represented as a random variable $X$ with probability distribution $P$ over the set $\mathcal X$ of possible events $x$ (events are elements or interactions). A standard measures of disorder is the \emph{variance}. The variance can be used for events that are numeric, such as the number of edges per node in a network, but not for types, such as species in a population. The variance of a random variable $X$, \begin{equation}
    \varX :=  \mathbb{E}[ (X - \mathbb{E}[X])^2]~,
\end{equation}  measures the average deviation from the mean. The equivalent notation in the physics literature is $\varX = \expX{(X - \expX{X} )^2}$. The broader a distribution of possible event values is the higher, in general, the variance. A second standard measure of disorder, the \emph{Shannon entropy}, is a function from information theory (see Appendix~\ref{sec.info-theory} for a brief introduction to information theory). The {Shannon entropy} of a random variable $X$ with probability distribution $P$ over events $x$ is defined as
\begin{equation}
H(X) :=  - \sum_{x\in {\mathcal X}} P(x) \log P(x)~.    
\end{equation}
The Shannon entropy measures the amount of uncertainty in the probability distribution $P$. In the case of all probabilities being equal, the distribution is a so-called uniform distribution. In this case all events are equally likely, and the uncertainty, and hence the Shannon entropy, over events is maximal. The Shannon entropy is zero when one probability is one and the others are zero. If, for example, the events $x$ were the possible outcomes of an election, then $H(X)$ would quantify the difficulty in predicting the actual outcome.

To illustrate these measures of disorder consider a network, such as the World Wide Web or a neural network. The disorder relevant to a network is structural disorder. A network with many nodes and edges between every pair of nodes is considered a network with no disorder. The origin of a given network structure is often studied with network-formation models. For an overview of this and other network-formation models see, for example,  \cite{newman_networks:_2010}. One of the first network-formation models is the so-called Erd\"os-Renyi random graph model (or just random graph model, Poisson model, or Bernoulli random graph) \cite{erdos1960evolution}.  
The Erd\"os-Renyi model is parametrised by the number of nodes $n$, the maximum number of edges $M$, and  a  parameter $p$ which is the probability of an edge being created between two existing nodes. Initially, the network has $n$ nodes and no edges. In a subsequent formation process, with probability $p$, two nodes are connected by an edge.  When $p=0$, the resulting network after many repetitions is a set of nodes without any edges. For $p=1$, the result is a highly connected network. 
For $p$ somewhere in between $0$ and $1$, the formation process yields a network with links between some nodes and some nodes having more links than others. In this case, the probabilistic nature of the link formation results in a disordered structure of the network. Hence, the disorder of the formation process is taken as a proxy for the disorder of the final network structure. Several properties of the fully formed network, such as the average path length and the average number of edges per node, can be expressed as functions of $n$, $M$, and $p$ only. These regularities emerge out of the disorder in the formation process.

The variance can be used to quantify the disorder of the network-formation process after assigning numeric values to the events `edge' and `no edge' -- for example $1$ and $0$, respectively. The variance of the binary probability distribution $P = \{p, 1-p\}$ of the Erd\"os-Renyi random graph model is $\varX = p(1-p)$, which is maximal for $p=1/2$. The Shannon entropy of the network-formation process can be computed without assigning numerical values to the events. The Shannon entropy of the binary probability distribution $P = \{p, 1-p\}$ of the Erd\"os-Renyi random graph model is  $H(X) = -p\log p - (1-p)\log(1-p)$, which is also maximal for $p=1/2$. Both measures are zero when $p=0$ and, due to symmetry, when $p=1$. If one were to measure the disorder in the final network structure itself, the variance and the Shannon entropy should be computed from the probability distribution over the node degrees. The result would be equivalent to the previous one in the sense that the degree distribution is trivial for $p=0$ (completely disconnected) and $p=1$ (fully or nearly fully connected), in which case both measures yield the value $0$% or close to $0$
. For non-trivial network structures both measures are non-zero. It was remarked above that a measure of order can be used as a measure of lack of disorder and vice versa. Hence, any of the existing measures of network structure, such as average path length or  clustering, can be used to measure disorder by monitoring their change. This approach to measuring disorder has been used in the study of Alzheimer's disease and its effect on neural connectivity in the human brain  (see \cite{bullmore2012economy} and references therein).

Temporal disorder in a sequence of events, such as the sequence of daily share prices on a stock market, is described with the language of stochastic processes. Disorder in a stochastic process is the lack of correlations between past and future events. 
A stochastic process is defined as a sequence of random variables $X_t$ ordered in time (see Appendix~\ref{sec.probtheory} for more details). Disorder is the lack of predictability of future events when past events are known. To quantify disorder in a sequence $X_{1}X_{2} \dots X_n$, the joint probability over two or more of the random variables is required, written as $P(X_{1}X_{2} \dots X_n)$. This is the probability of the events occurring together (jointly). When a joint probability of two events is known, then, in addition to their individual probability, it is known how likely they are to occur together. 
An example is the probability of certain genetic mutations being present and the probability of two mutations being present in the same genome. The joint Shannon entropy $H(X_{1}X_{2} \dots X_n  )$ over this distribution, 
\begin{align}
\begin{split}
H(X_{1}X_{2} \dots X_n  ) := - \sum_{x^n\in {\mathcal X}^n} & P(x_1 x_2 \dots x_n) \\
 & \cdot \log P(x_1 x_2 \dots x_n)~,    
 \end{split}
\end{align}
captures the lack of correlations. A measure of average temporal disorder is the so-called \emph{Shannon entropy rate}, \begin{equation}
h_n:= \frac{1}{n} H(X_{1}X_{2} \dots X_n  )~.
\end{equation}
The Shannon entropy rate measures the uncertainty in the next event, $X_n$, given that all $n-1$ previous events $X_1 \dots X_{n-1}$ have been observed. The lower the entropy rate, the more correlations there are between past and future events and the more predictable the process is.

A fly's brain is an example of a complex system where temporal disorder has been measured experimentally. 
\cite{van1997reproducibility} recorded  spike trains of a motion-sensitive neuron in the fly's visual system. 
From repeated recordings of neural spike trains, they constructed a probability distribution $P(X_1 X_2 \dots X_k)$ over spike trains of some length $k$. From this probability distribution they computed the joint Shannon entropy $H(X_1 X_2 \dots X_k)$ and the entropy rate $\frac1k H(X_1 X_2 \dots X_k)$. They repeated the experiments after exposing the fly to the controlled external stimulus of a visual image 
and computed the Shannon entropy and entropy rate again. 
They interpreted the difference in the entropies between the two experiments as the reduction in disorder of the neural firing signal when a stimulus is present. 

One speaks of the `diversity' of species in an ecosystem or of diversity of stocks in an investment portfolio rather than `disorder'. In the language of diversity, the elements, species or stocks, are called `types'. The simplest measure of diversity is the number of types or the logarithm of that number. A more informative measure takes into account the frequency of each type, this being the number of individuals of each species in a habitat or the number of each stock in the portfolio. Treating such frequencies as probabilities, a random variable $X$ of types $\xcal$ can be constructed, and the Shannon entropy $H(X)$ is used as a measure of type diversity. In ecology, diversity is  measured using the \emph{entropy power}, $2^{H(X)}$ (if the entropy is computed using $\log$ base $2$ or $e^{H(X)}$ if the entropy is computed using $\log$ base $e$) \cite{jost2006entropy}. It behaves similar to the entropy itself but has a useful interpretation: the entropy power is the number of species in a hypothetical population in which all species are equally abundant and whose species distribution has the same Shannon entropy as the actual distribution. 
 If the types are numeric, such as the size of pups in an elephant seal colony \cite{fabiani2004extreme}, diversity can be measured using the variance. Often a normalised form of variance, the \emph{coefficient of variation}, is used:
\begin{equation}
\textnormal{cv} := \frac{\sqrt{\mathrm{Var} X}}{\expX{X}}~.
\end{equation}
The coefficient of variation is the square root of the variance (also known as the standard deviation) divided by the mean. Its behaviour is equivalent to that of the variance. Broader distributions, such as a larger range of pup sizes in an elephant seal colony, result in a higher coefficient of variation. However, it allows the comparison of distributions with the same variance but different means. A distribution with a variance of $10$ and a mean of $20$ might be considered more diverse than a distribution with a variance of $10$ and a mean of $1,000$. Their coefficient of variation would reflect this difference.  

Scott Page, in his book `Diversity and Complexity'  \cite{page_diversity_2010}, distinguishes between three kinds of diversity: diversity within a type, diversity across types, and diversity of community composition. All three are measured by the Shannon entropy. In fact, they differ only  in what constitutes an event in the definition of the random variable. Other measures of diversity are the so-called `distance' measures and `attribute' measures. Distance measures of diversity, such as the Weitzman Diversity, take into account not only the number of types, but also how much they differ from each other \cite{weitzman1992diversity} and therefor require a mathematical measure of distance. Attribute-diversity measures assign attributes to each type and numerically weigh the importance of each attribute. For example, to compute an attribute diversity of phenotypes more weight is put on traits with higher relevance for survival (see \cite{page_diversity_2010} for more details).

\section{Feedback}
\label{sec.feedback}

The interactions in complex systems are iterated so that there is feedback from previous interactions, in the sense that the parts of the system interact with their neighbours at later times depending on how they interacted with them at earlier times. And these interactions take place over a similar time scale to that of the dynamics of the system as a whole. There is no measure of feedback as such. Instead, the effects of feedback such as nonlinearity or structure formation are measured. Hence, the mathematical tools that are used to measure order and nonlinearity can also be indicators of feedback. 

A common way to study feedback is to construct a mathematical model with feedback built into it. If the model reproduces the observed dynamics well, this suggests the presence of feedback in the  system that is being modelled. An example is the dynamics of a population of predator and prey species such as foxes and rabbits. The  growth and decline of these species can be modelled by the Lotka-Volterra differential equation model. It describes the change over time in population size of two species, the prey $x$ and its predator $y$, using the four parameters $A$, $B$, $C$, and $D$. $A$ and $C$ are parameters for the speed of growth of the respective species.  $B$ and $D$ quantify the predation.  \label{box.LVmodel} 
 The change over time in population size, $\dot{x}$ and $\dot{y}$, is given by the two coupled equations
	\begin{align}
	\begin{split}
	  \dot{x} &= Ax - B xy~,\\
	  \dot{y} &= - Cy + Dxy~.
	  \end{split}
	\end{align}
The fact that $x$ and $y$ appear in both equations ensures that there is a feedback between the size of each population. If $B$ or $D$ are zero, there is no feedback. 

For certain values of the parameters $A$, $B$, $C$, and $D$ the number of individuals of each species oscillates. When the overabundance of predators reduces the number of prey to below the level needed to sustain the predator population but the resulting decline in the number of predators allows the prey to recover, a  cycle of growth and decline results. For such oscillations to happen the time scale of growth, captured by $A$ and $C$, needs to be similar to the time scale of predation, captured by $B$ and $D$. Oscillations in predator-prey populations is a classic example of feedback.

A widely used computational tool for studying feedback are so-called \emph{agent-based models}. These models are computational simulations of agents undergoing repeated interactions following simple rules. In such a  simulation a usually large set of agents is equipped with a small set of actions that each agent is allowed to execute and a small set of (usually simple) rules defining the interaction between the agents. In any given round of a simulation an agent and an action, or two agents and an interaction, are picked at random. If the action (interaction) is allowed, it is executed. An agent-based simulation usually consists of many thousands of such rounds. One of the first agent-based models was the sugarscape model, pioneered by the American epidemiologist Joshua Epstein and computational, social and political scientist Robert Axtell  \cite{epstein1996growingMIT}. The sugarscape model is a grid of cells, some of which contain `sugar'; the others contain nothing. Agents %, with slightly diverse attributes,
`move' on this landscape of cells and `eat' when they find a cell containing sugar. Even this very simple setup produces emergent phenomena such as the feedback effect of the-rich-get-richer.

Agent-based models are frequently used to study feedback in the coherent dynamics of animal groups \cite{couzin2002collective}. \cite{couzin_self-organized_2003} describe observations of army ants in Soberania National Park in Panama. Army ants make an excellent study case for collective phenomena since they are able to form large-scale traffic lanes to transport food and building material over long distances. They even form bridges out of ants to avoid `traffic congestion'. These collective phenomena are impossible without the presence of feedback. The authors set up an agent-based simulation with simple movement and interaction rules for individual ants. Feedback is built in as an ant's tendency to avoid collision with other ants and in its response to local pheromone concentration. The simulations reproduce the observed lane formations and the minimisation of congestion. Such a simulation is not to be confused with the measurement of actual feedback in a real system.

There are other notions of feedback in the literature on complex systems. The computational notion of feedback is to `feed back' the output of a computation as input into the same computation. In this way, the outcome of future computations depends on the outcome of previous computations. This kind of feedback is particularly important for those who view nature to be inherently computational \cite{lloyd2006programming, davies2014information}. On this view, any loop in the  computational representation of a natural system indicates the presence of feedback. Nobel Laureate  Paul Nurse made a similar point when presenting his computational view of the cell  \cite{nurse_life_2008}.

The above tools for analysing feedback have in common that they do not assign a number to the phenomenon, as is done in the case of disorder or diversity. Instead, in most practical applications feedback is a tunable interaction parameter of a model or an observable consequence of the interactions which are programmed into a model.

\section{Non-Equilibrium}
\label{sec.measuresnoneq}

Complex systems are open to the environment, and they are often driven by something external. Non-equilibrium physical systems are treated by the theories of non-equilibrium thermodynamics \cite{de2013non} and stochastic processes \cite{van1992stochastic}. Stochastic complex systems, such as chemical reaction systems, are often studied using the statistics of Markov chains. Consider a system represented by a set of states, $S$, through which the system evolves in discrete time steps. Let $\{P_{ij}\}$ be a matrix of time-independent probabilities of transitioning from state $i$ to state $j$, with $\sum_j P_{ij} = 1$,\footnote{Some scientific fields use the reverse order, $\{P_{ji}\}$.} for all $i\in S$. Let $\pi_i$ be the probability of being in state $i$. If there exists a probability distribution $\pi^*$ such that, for all $ j$, 
\begin{align} 	\pi_j^* = \sum_i P_{ij}\pi_i^*~,
\end{align}
it is called the invariant distribution. In a stochastic model of a system of chemical reactions, for example, the chemical composition is represented as a probability distribution, and chemical reactions are represented as stochastic transitions from one reactant to another. A system is in chemical equilibrium if the chemical composition is time-invariant. Reactions are still taking place in chemical equilibrium, but the depletion of one reactant is compensated by other transformations such that the overall concentrations remain largely unchanged. A general framework to model non-equilibrium stochastic dynamical systems is that of stochastic differential equations \cite{ikeda2014stochastic}.  

For systems for which a description in terms of chemistry or thermal physics is unhelpful, such as the brain or the World Wide Web, information theory is often used to describe the equivalent of a non-equilibrium state. Based on the probability distribution over the relevant state space, a measure related to the mutual information (see Section~\ref{sec.order}) quantifies the distance to an equilibrium state. Consider the  probability distribution $P$ over current state space  $\mathcal{X}$ and the corresponding equilibrium distribution $Q$. The amount of non-equilibrium is then quantified by the so-called Kullback-Leibler divergence (or relative entropy) $D(P || Q)$:
\begin{align}
  D(P || Q) = \sum_{x\in \mathcal{X}} P(x) \log \frac{P(x)}{Q(x)}~,
\end{align}
which is zero only if $P = Q$, in other words only if the system is in equilibrium. Examples where this has been used are the predictive brain model \cite{friston2010free} and stochastic time sequences \cite{still2012thermodynamics}. 

\section{Spontaneous Order and Self-Organisation}
\label{sec.order}

Perhaps the most fundamental idea in complexity science is that of order in a system's structure or behaviour that arises from the aggregate of a very large number of disordered and uncoordinated interactions between elements. Such \emph{self-organisation} can be quantified by measuring the order that results -- for example, the order in some data about the system. However, measures of order are not measures of self-organisation as such since they cannot determine how the order arose. This is because the order in a string of numbers is the same regardless of its source. Whether the order is produced spontaneously as a result of uncoordinated interactions in the system or whether it is the result of external control cannot be inferred from measuring the order without background knowledge about the system. For example, the orderly traffic lanes to and from food sources formed by an ant colony are considered the result of a self-organising process since there is no mechanism which centrally controls the ants' behaviour, while the orderly checkout lines in a supermarket are the result of a centrally managed control system. A high measure of order, even when self-organised, is not to be confused with a high level of complexity since order is but one aspect of complexity. However, the plethora of measures of order which are labelled as measures of complexity reflects the ubiquity of order in complex systems and explains the frequent use of order as a proxy for complexity. 

Complex systems can produce order in their environment. It is important to remember that the order produced by the system is different from the order in the system itself. For example, the order of hexagonal wax cells built by honey bees is order produced by the system, while division of labour in the hive is order in the system. The hexagonal honeycomb structures are a form of spatial correlation which can be quantified by correlation measures, some of which are discussed in the following.

A correlation function is a means to measure dependence between random variables; therefore, it is a statistical measure. The \emph{covariance} is a standard measure of correlation. 
For any two numeric random variables $X$ and $Y$,  the covariance, 
\begin{equation}
\label{eq.covMaintext}
\textnormal{cov}(X, Y) = \mathbb{E}[X Y] -  \mathbb{E}[X] \mathbb{E}[Y]~,
\end{equation}
is the difference between the product of the expectations and the expectation of the product. If the two random variables are uncorrelated, this difference is zero. 
From the covariance a dimensionless correlation measure is derived, the so-called \emph{Pearson correlation}. It is the most standard measure of correlation and defined as follows:

\begin{align}
\textnormal{corr}(X, Y) := \frac{\textnormal{cov}(X, Y)}{\sigma_{X} \sigma_{Y}}~,
\end{align}
where $\sigma$ is the square root of the variance, known as the standard deviation. 

A measure of correlation derived from information theory is the \emph{mutual information}. For two random variables $X$ and $Y$, the mutual information is a function of the Shannon entropy $H$ (see Section~\ref{sec.disorder.measure}):
\begin{equation}
    I(X;Y) = H(X) + H(Y) - H(X,Y)~.
\end{equation}

The mutual information measures the difference in uncertainty between the sum of the individual random variable distributions and the joint distribution. If there are any correlations between the two variables, the uncertainty in their joint distribution will be lower than the sum of the individual distributions. This is a mathematical version of the often repeated statement that `the whole is more than the sum of its parts'. If the whole is different from the sum of the parts, it means that there are correlations between the parts. For two completely independent random variables, on the other hand, $H(X) + H(Y) = H(X,Y)$ and the mutual information is zero.

An example of covariance as a measure of order is the study of bird flocking by William Bialek, Andrea Cavagna, and  colleagues \cite{bialek_statistical_2012}. 
They filmed flocks of starlings in the sky of Rome (containing thousands of starlings) and extracted the flight paths of the individual birds from these videos. Each bird's different flight directions over time were represented as a random variable, and the random variables of all birds were used to compute their pairwise covariances.\footnote{They used the convention from statistical mechanics in which the uncorrelated average product is not subtracted. Thus, their covariance is the statistical mechanical correlation function $\mathbb{E}[X Y]$.} This list of covariances was fed into a computer simulation that modelled the flock of birds as a condensed matter system, which is defined by the interaction between close-by `atoms' only. 
The computer simulation of such a very simple  model with pairwise interactions only and no further parameters, produced a self-organising system that very closely resembled the self-organising movement originally observed. 
A similar analysis was been done on network data of cultured cortical neurons, corroborating the idea that the brain is %local
self-organising \cite{schneidman_weak_2006}. 

The order in a flock of starlings is a spatial order persistent over time. Systems in which the focus is more on the temporal aspect of the order are neurons and their spiking sequences, for example, or the monsoon season and its patterns. Order in these systems is studied by representing them as sequences of random variables $X_1 X_2 \dots X_t$ with a joint probability distribution $P(X_1 X_2 \dots X_t)$. Such sequences we encountered above in the study of disorder.
Several authors, independently, introduced the mutual information between parts of a sequence of random variables as a measure of order in complex systems, under the names of \emph{effective measure complexity} (EMC) \cite{grassberger_toward_1986}, \emph{predictive information} ($\textnormal{I}_\text{pred}$) \cite{bialek_predictability_2001}, and \emph{excess entropy} (E) \cite{crutchfield_regularities_2003}.
 Consider the infinite sequence of random variables $X_{-t} X_{-t+1} \dots X_0 X_1 X_2 \dots X_t$, which is also called a stochastic process. 
The information theoretic measure $\textnormal{I}_\text{pred}$ (or EMC or E) of correlation between the two halves of a stochastic process  
  is defined as  the mutual information between  the two halves:
  \begin{align}\label{eq.Ipred}
   \textnormal{I}_\text{pred}   	&:= \lim_{t\to\infty} I(X_{-t} X_{-t+1} \dots X_{-1};X_{0} X_1 \dots X_{t})~.
 \end{align}
 There is, of course,  never an infinite time course of data, and the limit $t\to\infty$ is never taken in practice. 

\cite{palmer_predictive_2015} %Stephanie Palmer, William Bialek, and colleagues
measured the predictive information in retinal ganglion cells of salamanders. Ganglion cells are a type of neuron located near the inner surface of the retina of the eye. In the lab, the salamanders were exposed, alternatively, to videos of natural scenes and to a video of random flickering. While a video was showing, the researchers recorded a salamander's neural firings. 
Repeated experiments allowed them to infer the joint probability distribution $P(X_{-t} \dots X_t)$ over the ganglion cell firing rates and to compute the predictive information {I}$_\text{pred}$ contained in it. They found that I$_\text{pred}$ was highest when a salamander was exposed to naturalistic videos of underwater scenes. This shows that the order in the natural scenes is reflected in the order of the neural spike sequences. The authors also think that it shows the neural system not only responds to a visual stimulus, but also makes predictions about it. 

Quantifying predictability and actually predicting what a system is going to do are, of course, two different things. In order to make a prediction one first has to have a model, for example, inferred from a set of measured data.

\section{Nonlinearity}
\label{sec.nonlinearity}

There are several different phenomena addressed with the same label of `nonlinearity'. Each phenomenon requires its  own measure. Power laws are probably the most prominent examples of nonlinearity in complex systems. But correlations as a form of nonlinearity are equally important, and these two are not completely separate phenomena.

\subsection{Nonlinearity as Power Laws}

A power law is a relation between two variables, $x$ and $y$, such that $y$ is a function of the power of $x$ -- for example, $y = x^\mu$. Quite a few phenomena in complex systems, such as the relation between metabolism and body mass or the number of taxpayers with a certain income and the amount of this income, follow a power law to some extent.  The power law of metabolism for mammals was first discussed by Max \cite{kleiber1932body} in 1932. It is now well established that, to a surprising accuracy, the metabolic rate of mammals, $R$, is proportional to their body mass, $m$, to the power of $3/4$: $R \propto m^{3/4}$ (see \cite{west_general_1997} and references therein).  Because $3/4$ is less than $1$, a mammal's metabolism is more efficient the bigger the mammal; an elephant requires less energy per unit mass than a mouse. This is a nonlinear effect since doubling the body size does not double the energy requirements. It is also another instance of the often repeated, but confused, statement that, in complex systems, the whole is more than the sum of its parts. The whole is never more than the sum of its parts when interactions are taken into account. 

The relation between taxpayer bracket and number of people in this bracket is an instance of a statistical distribution that exhibits a power-law behaviour. 
Other examples of statistical distributions with a power-law behaviour are the number of metropolitan areas relative to their population size, the number of websites relative to the number of other websites linking to them, and the number of proteins relative to the number of other proteins that they interact with (for reviews, see \cite{newman_power_2005, sornette_critical_2009}). 

Statistical distributions with a power-law behaviour are defined in terms of random variables. 
Consider a discrete random variable $X$ with positive events $x >0$ and probability distribution $P$. The distribution $P$ follows a power law if 
\begin{align}\label{eq.powerlaw}
P(x) = c x^{-\gamma}~,
\end{align}
for some constant  $\gamma > 1$ and normalisation constant $c = (\gamma - 1)/(x_{\text{min}}^{1-\gamma})$, where $x_{\text{min}}$ is the smallest of the $x$ values. A cumulative distribution with a power-law form is also called a  Pareto distribution; a discrete distribution with a power-law form is also called a Zipf distribution (for a review, see  \citealt{mitzenmacher_brief_2004}).  Eq.~\ref{eq.powerlaw} can be written as $\log P(x) = \log c -\gamma\log x$, which says that plotting $\log P(x)$ versus $\log x$ yields a straight line with  slope $-\gamma$. 
Therefore, the presence of a power law in real-world distributions is  often determined by fitting a straight line to a log-log plot of the data. Although this is common practice, there are many problems with this method of identifying a power-law distribution \cite{clauset_power-law_2009}.

A power-law distribution has a well-defined mean for $\gamma \leq 1$ over $x\in [1,\infty)$ and a well-defined variance for $\gamma \leq 2$. Power-law distributions are members of the larger family of so-called fat-tailed distributions. These probability distributions are distinct from the most common distributions, such as the Gaussian or normal distribution, in that events far away from the mean have non-negligible probability. Such rare events have obtained the name `black swan' events;  they come as a surprise but have major consequences \cite{taleb2007black}.

\subsection{Nonlinearity versus Chaos}

Nonlinearity in complex systems is not to be confused with  nonlinearity in dynamical systems. Nonlinear dynamical systems are sets of equations, often deterministic, describing a trajectory in phase space, either continuous or discrete in time. Some of these systems exhibit chaos, which is the mathematical phenomenon of extreme sensitivity of the trajectory on initial conditions. An example of a discrete dynamical system exhibiting chaos is the logistic map. The logistic map, $x_{t+1} = rx_t(1 - x_t)$ where $t$ indexes time, is a simple model of population dynamics of a single species, as opposed to two species, discussed above in the context of feedback. This map is now a canonical example of chaos. 

Actual physical systems studied by dynamical systems theory, such as a chaotic pendulum, need not have any of the features of complex systems. Certainly, chaos and complexity are two distinct  phenomena. On the other hand, the time evolution of many complex systems is described by nonlinear equations. Some climate dynamics, for example, are modelled using the deterministic Navier-Stokes equations, which are a set of nonlinear equations describing the motion of fluids. Another example of a nonlinear equation used to describe many complex systems is the Fisher-KPP differential equation \cite{fisher_wave_1937}. Originally introduced in the context of population dynamics, its application ranges from plasma physics to physiology and ecology.

\subsection{Nonlinearity as Correlations or Feedback}
\label{subsec.nonlinear}

For some the notion of nonlinearity in complex systems is synonymous with the presence of correlations  (for instance, 
\cite{mackay_nonlinearity_2008}).  If two random variables $X$ and $Y$ are independent, their joint probability distribution $P(XY)$ is equal to the product distribution $P(X) P(Y)$. When this equality does not hold, then there must be correlations between $X$ and $Y$.

Defining `nonlinearity' in terms of the presence of correlations is not to be confused with  linear versus nonlinear correlations. In the language of statistical science, two variables $X$ and $Y$ are linearly correlated if one can be expressed as a scaled version of the other,  $X = a + cY$, for some constants $a$ and $c$. The Pearson correlation coefficient, for example, detects linear correlations only. 
The mutual information, on the other hand, detects all correlations, linear as well as nonlinear. 

To others, mainly social scientists, `nonlinearity' means that the causal links of the system form something more complicated than a single chain. A system with causal loops, indicating feedback, would count as `nonlinear' in this view 
\cite{blalock1985causal}. 

The different definitions of nonlinearity discussed here are all ubiquitous in complex systems research, so it is not surprising that nonlinearity is often mentioned as essential to complex systems.

\section{Robustness}
\label{sec.robustness}

Several phenomena  are often grouped together under the umbrella of `robustness'. A system might be robust against perturbation in the sense of maintaining its structure or its function upon perturbation, which some refer to as `stability'. Alternatively, a system might be robust in the sense that it is able to recover from a perturbation; this is also called `resilience'.

Strictly speaking, robustness is the property of a model, an algorithm, or an experiment that is robust against the change of parameters, of input, or of assumptions. But usually, in the context of complex systems, 
robustness refers to the stability of structure, dynamics or behaviour in the presence of perturbation. All order and organisation must be robust to some extent to be worth studying. 
Several tools are available for studying robustness; the most frequently used are tools from dynamical systems theory and from the theory of phase transitions. Brief descriptions are given outlining the role of these tools in the study of complex systems.

\subsection{Stability Analysis}

The system of predator and prey species sharing a habitat, which was discussed above (see the Lotka-Volterra population model in Section~\ref{sec.feedback} and Section~\ref{sec.nonlinearity}), is an example of a stable dynamical system. After some time the proportion of the two species becomes either constant or oscillates regularly, independent of the exact proportion of species in the beginning. 
A dynamical system is called `stable' if it reaches the same equilibrium state under different initial conditions or if it returns to the same equilibrium state after a small  perturbation. Stability analysis is prevalent in physics, nonlinear dynamics, chemistry, and ecology. A reversible chemical reaction, for example, might be stable with respect to forced decrease or increase of a reactant, which means the proportion of reactants and products returns to the same value as before the perturbation. 
Other examples of complex systems which are represented as dynamical systems are food webs with more than two species \cite{pimm1982food, rooney2006structural},  genetic regulatory networks \cite{de_jong_modeling_2002}, and neural brain regions \cite{friston2009causal%, stephan_dynamic_2006
  }.

 For any given dynamical system described by a state vector ${\bf x}$ and a set of (possibly coupled) differential equations $dx_i/dt = f_i({\bf x})$, a stable point, a so-called fixed point, is a solution to the equations $dx_i/dt = 0$. Stability analysis classifies these fixed points  into stable and unstable ones (or possibly stable in one direction and unstable in another). Assuming the system is at one of its fixed points, the effect of a small perturbation on the system's dynamics is found by analysing the Jacobian matrix $J$, a linearisation of the system, which is defined as 
 \begin{align}
\left[J_{ij}\right] = \left[\frac{\partial f_i }{\partial x_j}\right]~.
\end{align}
  If the eigenvalues of the Jacobian evaluated at a given fixed point  all have real parts that are negative, then this point is a stable fixed point and the system returns to the steady state upon perturbation. If any eigenvalue has a real part that is positive, then the fixed point is unstable and the system will move away from the fixed point in the direction of the corresponding eigenvector, usually towards another, stable, fixed point. For an introduction to fixed-point analysis of dynamical systems, see, for example, \cite{strogatz2014nonlinear}. 
 Any stable fixed point is embedded in a so-called basin of attraction. The size of this basin quantifies the strength of the perturbation which the system can withstand and, therefore, is a measure of the  stability of the system at the fixed point \cite{strogatz2014nonlinear}. 
 Stability analysis is widely used in  ecology  \cite{holling1973resilience,scheffer_complex_2010}.

  Viability theory combines stability analysis of deterministic dynamical systems theory with control theory \cite{aubin1990survey, aubin2009viability}. It  extends stability analysis to more general, non-deterministic systems and provides a mathematical framework for predicting the effect of controlled actions on the dynamics of such systems, with the aim of regulating them. Viability theory has been applied to the resilience of social-ecological systems \cite{bene2018resistance, deffuant2011viability}. 
  
  A similar, though mostly qualitative, use of the ideas of stability and viability is found in the analysis of \emph{tipping points} in climate and ecosystems. Tipping points are the points of transition from one stable basin of attraction to another, instigated by external perturbations \cite{scheffer_complex_2010}.

\subsection{Critical Slowing Down and Tipping Points}

The time it takes for a system to return to a steady state after a perturbation is a stability indicator complementary to the fixed-point classification and the size of the attractor basin. The longer it takes the system to recover after a perturbation the more fragile the system is. An increase in relaxation time can indicate a \emph{critical slowing down} and the vicinity to a so-called  tipping point or phase transition. When a system is close to a tipping point, it does not recover anymore from even very small perturbations and moves to a different steady state which is possibly very far away from its previous state. Finding measurable indicators for nearby tipping points has been of considerable interest, in particular since ecological and climate systems have begun to be characterised by stability analysis and their fragility is being recognised more and more  \cite{scheffer_complex_2010, scheffer2015generic}.

Mathematically, the vicinity to a tipping point is recognised by the functional dependency of the recovery time on the perturbation strength. A system which is close to a tipping point exhibits a recovery time that grows proportional to perturbation strength to some power. This  \emph{scaling law}, associated with critical slowing down, is a well-known phenomenon  in the statistical mechanics of phase transitions. 
The standard example of a phase transition in physics is the magnetisation of a material as a function of temperature. The  magnetisation density $m$ is proportional to the power of the temperature difference to a critical temperature, $m \propto |T - T_C|^{-\alpha}$. $T_C$ is the {critical temperature}, the equivalent to a tipping point, at which the magnetisation diverges and the system undergoes a phase transition.

Another signature of a nearby tipping point is an increase in fluctuations. In general, a perturbed system fluctuates around a steady state  before settling back down. The larger the  
length or time scale on which the fluctuations are correlated, the closer the system is to a tipping point. 

Experimentally, one might expose a system to increasingly strong perturbations and measure the time it takes the system to come back to its steady state. Such measurements yield the response of the system as a random variable $S$ as a function of spatial coordinate ${\bf x}$ and time $t$. The covariance $\textnormal{cov}(S({\bf x}, t),S({\bf x}+{\bf r}, t+\tau))$ (see Section~\ref{sec.order}) between the random variable at time $t$ and spatial location ${\bf x}$ and the same variable at some later time $t+\tau$ and some displaced location ${\bf x} + {\bf r}$ is a measure of the temporal and spatial correlations in time. The equivalent measure in physics is the so-called \emph{auto-correlation function}, denoted by $C({\bf r}, \tau)$, defined, in physics notation, as 
\begin{align}
C({\bf r}, \tau) = \expX{S({\bf x}, t) S({\bf x}+\bf{r}, t + \tau)}~.
\end{align}
It differs from the covariance by not subtracting the product of the marginal expectations, $\expX{S({\bf x}, t)} \expX{S({\bf x}+\bf{r}, t + \tau)}$ (in statistics notation, $\mathbb{E}[S({\bf x}, t)] \mathbb{E}[S({\bf x}+\bf{r}, t + \tau)]$), from the expectation of the product (cf. eq.~\ref{eq.covMaintext}). 

When correlations decay exponentially in time, $C$ is proportional to $e^{-k\tau}$, where $k$ is an inverse time. After time $\tau = 1/k$ correlations have decayed to a fraction $1/e$ of the value they had at time $t$, and $\tau=1/k $ is the so-called characteristic time scale. Equally, when correlations decay exponentially with distance, the distance $|\bf r|$ at which they have decayed to a fraction $1/e$ of the value at $|{\bf r}| = 0$ is the characteristic length scale.
Critical slowing down is accompanied by %long-range
fluctuations that decay slower than exponentially. The signature in the auto-correlation function $C$ is a power-law decay either in time or in space, $C \propto |{\bf r}|^{-\alpha}$ or $C \propto \tau^{-\alpha}$.  Theoretically, at the point of a phase transition the correlation length becomes infinite. At that point the system has correlations on all scales and no characteristic length nor time scale anymore. %, depending on whether the dynamic or the structure is undergoing a phase transition. 
%, there is no characteristic length or time scale. 
A correlation length which captures nonlinear correlations has been based on the mutual information \cite{dunleavy2015mutual}.

An example of a complex system where critical slowing down has been measured is a population of cyanobacteria under increasing irradiation. The bacteria require light for photosynthesis, but irradiation levels that are too high are lethal. For protection against destructively high irradiation levels, bacteria have evolved a shielding mechanism. Annelies Veraart and colleagues 
exposed cell cultures of cyanobacteria to varying intensities of  irradiation and studied the subsequent shielding process \cite{veraart_recovery_2012}. When the irradiation was relatively weak the bacterial population quickly recovered after enacting the mutual shielding mechanism by which the bacteria protect each other. The stronger the radiation, the longer it took the population to build up the necessary shielding and recover afterwards. Veraart and her colleagues measured a critical slowing down with a power-law-like behaviour.  Once the light stress reached a certain threshold,  equivalent to a critical point, the population collapsed. The new steady state that the population had tipped into was that of death.  
There are many other complex systems where critical slowing down has been suspected or observed -- for example, in the food web of a lake after introduction of a predator species \cite{carpenter2011early}, in marine ecosystems in the Mediterranean after experimental removal of the algal canopy  \cite{benedetti2015experimental}, and in paleoclimate data around the time of abrupt climatic shifts \cite{dakos2008slowing}. For a review of critical slowing down in ecosystems, see \cite{scheffer2015generic}. 
For many more examples of criticality in complex systems, ranging from geological to financial systems, see  \cite{sornette_critical_2009}.

\subsection{Self-Organised Criticality and Scale Invariance}

Power laws are an example of nonlinearity, as discussed in Section~\ref{sec.nonlinearity}. Power-law behaviour is also an example of instability since a power-law behaviour in the recovery time is the signature of a system being driven towards a critical point, as discussed. 
It is, therefore, unexpected that many complex systems exhibit a power-law behaviour without any visible driving force and that they are nevertheless relatively stable. It appears that such systems stay close to a critical point `by their own choice', a phenomenon called \emph{self-organised criticality}. When it was discovered in a one-dimensional lattice of coupled maps \cite{keeler1986robust} and later observed in a computer model of avalanches  \cite{bak1988self}, it  sparked a whole surge of studies into the mechanism behind self-organised criticality. 
This surge was fueled by experimental observations of power-law-like behaviour in a range of different systems, such as the Earth's mantle and 
the magnitude and timing of earthquakes and their afterquakes, or the brain and the timing of neurons  \cite{zoller_observation_2001, sornette_predictability_2002, bullmore_complex_2009}. In these systems, the relevant observable, magnitude or timing, was measured as a histogram of frequencies of events. The  probability distribution $P(x)$ of events $x$, constructed from the data, decays approximately as a power law, $P(x) = c x^{-\gamma}$. As remarked above, the true functional form of these decays is still debated; it is rarely more than an approximate power law \cite{clauset_power-law_2009}. 
A power law implies so-called \emph{scale invariance}, since ratios are invariant to scaling of the argument: 
$
P(cx_1) / P(cx_2) = P(x_1) / P(x_2)$. 
Scale invariance has been observed in many natural as well as social complex systems \cite{sornette_critical_2009}, including scale invariance of the statistics of population sizes of cities  \cite{bettencourt2013origins}. 

While a power law in the auto-correlation function indicates instability and the vicinity of a critical point, a power law in a statistical distribution may indicate self-organised criticality which is associated with stability. 
Three decades after the discovery of self-organised criticality, there still is no known mechanism for it. The seeming contradiction between the robustness of a complex system, one of its emerging features, and the inherent instability of systems close to a tipping point remains unresolved. 
 For a review of self-organised criticality, see  \cite{pruessner_self-organised_2012} and \cite{watkins201625}. 

\subsection{Robustness of Complex Networks}

Network structures are ubiquitous in the interactions within a complex system. It is therefore not surprising that complex networks have grown into their own subfield of complex systems research. Many examples of networks have been mentioned in this book, from protein-protein interactions and neural networks to financial networks and online social networks. 
A network is a collection of nodes connected via edges. The \emph{degree} of a node is the number of edges connected to it. The nature of nodes and edges differs for each system. In protein-protein networks the nodes are proteins;  two nodes are connected by an edge if they interact, either biochemically or through electrostatic forces. 
A \emph{path} is a sequence of nodes such that every two consecutive nodes in the sequence are connected by an edge. The \emph{ path length} is the number of edges traversed along the sequence of a path. 
%The \textbf{shortest path} between two nodes is the sequence with the minimum number of traversed edges to get from one node to the other. 
The average shortest path is the sum of all shortest path lengths divided by their number. 
%The \emph{diameter} of a network is the longest of the shortest path between any two nodes.
The phrase `six degrees of separation' refers to the average path length between nodes in social networks. This goes back to a now famous experiment performed by Stanley Milgram and his team in the 1960s \cite{milgram1967small}. Milgram gave letters to participants randomly chosen from the population of the United States. The letters were addressed to a person unknown to them, and they were tasked with handing their letter to a person they knew by first name and who they believed would be more likely to know the recipient. This led to a chain of passings-on  for each letter. Surprisingly, letters reached the addressee, on average, after only five intermediaries. The stability of average path length is one proxy for the robustness of a network. When edges or nodes are removed from the network and the average path length stays more or less the same, the network is considered robust (in this respect). 
%Complex networks exhibit their own form of robustness that is structural rather than dynamic. A simple measure of a network's robustness is its density of links. Two further structural properties of interest in a network are its diameter and the average shortest path. A typical perturbation is the removal of a node or a link.
 Reka Albert, Hawoong Jeong and Albert-L\'asl\'o Barab\'asi  found that the Internet and the World Wide Web are very robust in precisely this way \cite{albert_error_2000}. The shortest path is hardly affected upon the random removal of nodes. Albert and her colleagues studied the structure of the World Wide Web and the Internet by taking real-world data and artificially removing nodes in a computer simulation. Plotting the shortest path  against the fraction of nodes removed from the network revealed that the path length initially stayed approximately constant. Only once a large fraction of the nodes had been removed did the length suddenly and dramatically increase. This sudden increase is a form of phase transition between a well-connected phase and a disconnected phase. It is seen already in the simplest model of networks, the Erd\"os-Renyi random graph model discussed above (see  \citealt{newman_networks:_2010}). Other real-world networks exhibiting this structural form of robustness are protein networks \cite{jeong2001lethality}, food webs \cite{dunne2002food}, and social networks \cite{newman2002email}. Robustness is always with respect to a feature or function. Robustness with respect to one feature might not imply robustness with respect to another. The Internet, for example, is robust against random removal of nodes (servers), but it is considerably less robust to targeted removal of the highest-degree nodes.

%Measures of robustness often do not assign a number but provide tools for studying robustness, similar to the tools used to measure feedback.

\section{Nested Structure and Modularity}
 \label{sec.modularity}

Nested structure and modularity are two distinct phenomena, but they may be related. `Nested structure' refers to structure on multiple scales. `Modularity' is a functional division of labour, or specialisation of function among parts, or a structural modularity and frequently all of these together.

Structural modularity is a property much discussed especially in the context of networks, where it is referred to as `clustering'. A cluster in a network is a  collection of nodes that have many edges between one another compared to only few edges to nodes in the rest of the network. A simple example is the network of online social connections such as the network of `friends' on Facebook. This network of social connections tends to be highly clustered since two `friends' of any given user are more likely to also be `friends' than to be unrelated. 

Finding clusters in networks has received considerable attention, and many so-called clustering algorithms have been proposed. For an introduction to clustering algorithms, see, for example, \cite{newman_networks:_2010}. 
All clustering algorithms follow a similar principle. Given a network, they initially  group the nodes into arbitrary communities, and, by some measure unique to each technique, they quantify the linking strength within each community and that in between communities. Information theoretic distance is one such measure \cite{rosvall_maps_2008}. The algorithms then optimise the communities by moving nodes between them until the linking strength within each community is maximised and the linking strength in between communities is minimised. There is usually no unique solution to this optimisation problem, and the identified clusters might differ from algorithm to algorithm. The presence of clusters alone is not sufficient for modularity since the network could consist of one gigantic cluster, with every node being connected to most other nodes, and have no modularity at all. 

Once a community structure of a network has been identified, the extent to which it is modular can then be quantified. One of the first measures designed to quantify structural modularity is the \emph{modularity} measure by Mark Newman and Michelle Girvan 
\cite{newman2004finding}. It assumes that a community structure of a given network has been identified and that  $k$ clusters of nodes have been found. From these $k$ clusters, a $k \times k$ matrix $\textbf{e}$ is constructed in which the entries $e_{ij}$ are the fraction of edges that link nodes in cluster $i$ to nodes in cluster $j$. The matrix entries can be interpreted as the joint probabilities $\Pr(i,j)$ for the event of an edge to be attached to a node in cluster $i$ and the joint event of this edge to end on a node in cluster $j$. If these two events are independent, the joint probability distribution is equal to its product distribution, $\Pr(i,j) = \Pr(i)\cdot \Pr(j)$. If, on the other hand, $\Pr(i,j) \neq \Pr(i)\cdot \Pr(j)$, then the probability $\Pr(i,j)$ is dependent on whether $i$ and $j$ are the same cluster ($i=j$) or not. With such a dependence present, there is modularity in the network. This condition of a joint probability distribution being a non-product distribution was a condition for `nonlinearity as correlations' (Section~\ref{subsec.nonlinear}). 

Newman and Girvan use this dependency condition to define modularity $Q$  as any deviation of the joint probability distribution $\Pr(i,i)$ of edges connecting nodes within the same cluster from the product distribution $\Pr(i)\cdot \Pr(i)$. In this sense, modularity is a form of nonlinearity as correlations. In the above matrix notation, the probability $\Pr(i) = \sum_{j} e_{ij}$. This can be understood as the so-called marginal probability of picking any edge in the network and for that edge to start in cluster $i$. Modularity is then defined as:
\begin{equation}
Q := \sum_i \left( e_{ii} - \left( \sum_j e_{ij} \right)^2 \right)~.
\end{equation}
This measure of modularity is also taken as an optimisation function for community detection algorithms, but limitations to its effectiveness have been pointed out  \cite{fortunato2007resolution, brandes2007modularity}. 

Many natural systems exhibit  structure that is repeated again on a smaller scale; the structure is nested within itself. A cauliflower exhibits this particular form of spatial scale invariance in the structure of the florets consisting of smaller florets and so forth. Beno\^it Mandelbrot discovered the mathematics of such nested structures, for which he coined the term \emph{fractal}. Fractals are mathematical objects with a perfect  scale invariance, a repetition of structure at an infinite number of scales \cite{falconer2004fractal}.  Mandelbrot's now famous book, \emph{The Fractal Geometry of Nature} \cite{mandelbrot1983fractal}, revealed the ubiquitous presence of fractal structure in natural systems, both living and nonliving. Fractals have the mathematical property of a non-integer dimension, and therefore fractal dimension is sometimes used as an indicator of nested structure (e.g., in ecology;  \cite{sugihara1990applications}). For example, a circle has dimension $2$, a sphere has dimension $3$, and the dimension of a cauliflower is estimated at $2.8$ \cite{kim2004fractal}.

Another indicator for multiple scales is the power-law decay in a correlation function (see Section~\ref{sec.robustness}).  For example, the number of websites in the visible World Wide Web as a function of their degree approximately follows a power law with an exponent $\gamma$, which, in 1999, was estimated at $2.1$ \cite{barabasi1999emergence}. This power-law decay is due to clusters of websites being nested within bigger clusters of websites. The World Wide Web has tens of billions of web pages, but only a few dozen domains own most of the links.\footnote{A domain is the name you need to buy or register, such as {google.com}. Any web page with a url within this domain name, such as {www.google.com/maps} is part of this domain.} These central domains are linked to each other, as well as to web pages within their own domain, and they also connect to large clusters of less-well-connected domains. Each of these clusters has, again, a few highly connected domains. This structure of clusters of sites with a few highly linked domains repeats at ever smaller scales. This self-similar nesting of clusters is much studied in complex networks  \cite{newman_structure_2003, ravasz_hierarchical_2003}. Methods based on statistical inference for identifying  nested clusters have also been developed \cite{clauset_hierarchical_2008}.

The presence of scale invariance in the degree distribution of a network can be reproduced by a model of network growth first considered by Derek %de Solla Price 
\cite{price1976general}. Starting with a small network, new nodes are added and connected by an edge to an existing node with a probability proportional to the existing node's degree. Hence, any new edge will affect the probability of future edges being added. Connecting a large number of nodes following this rule results in a network where a few nodes have a very high number of edges, and most nodes have very few. The algorithm is called the \emph{preferential attachment algorithm}. It is a variant of a random graph model (see Section~\ref{sec.disorder.measure}), and it describes the rich-get-richer effect seen in economics. It clearly has feedback built into it. The initial degree distribution might be uniformly random, but, after many iterations, it gets locked into a very skewed distribution due to the feedback of previously formed edges on future edge formation. 
The preferential attachment mechanism illustrates why power laws have been, and still are, a central theme in many studies of complex systems. Power-law-like behaviour can serve as an indicator for several of the features of complex systems identified in this book: nonlinearity, (lack of) robustness, nested structure, and feedback. This also suggests that these phenomena are not isolated from each other. 

\section{History and Memory}

The various measures of complexity measure different features of complex systems, all of which arise because of their histories. Hence, other measures can be used as proxies for history. For example, a network may have a definite growth rate so that the size of the network can be used as a measure of its age. Another way to measure the history of a complex system is to measure the structure it has left behind, because the more of it there is, the longer the history required for it to spontaneously arise as a result of the complex system's internal dynamics and interaction with the environment. However, in some cases, the structure in the world is built very quickly and deliberately rather than arising spontaneously, like a beaver's dam or a ploughed field. Background knowledge is needed to know how to relate such structure to history. There are no direct measures of history used in practice, but the logical depth discussed in the next section was introduced to capture the idea that complex systems require a long history to develop. Any measure of correlations in time, including the statistical complexity discussed below, can be considered a measure of memory.

\section{Computational Measures}
\label{sec.comp-measures}

Many of the growing number of measures of complexity are based on computational concepts such as algorithmic complexity and compressibility.\footnote{\cite{lloyd_measures_2001} produced a `non-exhaustive list' of over forty, and many more measures have been defined since then.} The previous section showed that complexity measures capture features of complexity but not complexity as such. This section discusses measures that consider complex systems to be computational devices with memory and computational power. All of these measures are reminiscent of thought experiments in that they are not implementable in practice or even in principle. Although these measures are now decades old (and none measure complexity as such) they are included here because they have had a considerable influence on thinking about complexity. We explain what feature of complexity each measures.

\subsection{Thermodynamic Depth}
\label{sec.th.depth}

\emph{Thermodynamic depth} was introduced by physicists Seth Lloyd and Heinz Pagels  \cite{lloyd_complexity_1988}. Lloyd and Pagels started out with the intuition that a complex system is neither perfectly ordered nor perfectly random and that a complex system plus a copy of it is not much more complex than one system alone. To specify the order of a complex system they consider the physical state of the system at time $t_n$, calling it $s_n$. In any stochastic setting, a given state can be preceded by more than one state. In other words, the set of states a system was in at times $t_1$ to $t_{n-1}$, a trajectory of length $n-1$, is not unique. Assigning a probability to any such trajectory which leads to state $s_n$, $\Pr(s_1, s_2, \dots, s_{n-1} | s_n)$, the {thermodynamic depth} of state $s_n$ is  defined as $-k \ln \Pr\left(s_1, s_2, \dots,s_{n-1} | s_{n}\right)$ averaged over all possible trajectories $s_1, s_2, \dots,s_{n-1}$,
\begin{align}
\begin{split}
 \mathcal{D}\left(s_n\right)  = -k \sum_{s_1,\dots,s_{n-1}} & \Pr\left(s_1, s_2, \dots, s_{n-1} | s_n\right) \\
 & \cdot \ln \Pr\left(s_1, s_2, \dots, s_{n-1} | s_n\right)~,
 \end{split}
\end{align}

where $k$ is the Boltzmann constant from statistical mechanics. In this view, the complexity of a system is given by the thermodynamic depth of its state. The intuition that the thermodynamic depth is intended to capture is that systems with many possible and long histories are more complex than systems which have short, and thus necessarily fewer possible, histories. What this definition leaves open, and arguably subjective, is how to find the possible histories, their lengths and what probabilities to assign to them \cite{ladyman2013complex}. Thus, practically, the measure is not implementable.
The rate of increase of thermodynamic depth when considering histories further and further back in time is mathematically an entropy rate, which is a measure of disorder (see Section \ref{sec.disorder.measure}). Thus, while the intention was for thermodynamic depth to be a measure of history, it is in fact a measure of disorder. This was pointed out in \cite{crutchfield_thermodynamic_1999}. 

\subsection{Statistical Complexity and True Measure \newline {Complexity}}

The \emph{quantitative theory of self-generated complexity}, introduced by physicist Peter Grassberger  \cite{grassberger_toward_1986}, and \emph{computational mechanics}, introduced by physicists James Crutchfield and Karl Young   \cite{crutchfield_inferring_1989,crutchfield_calculi_1994}, are similar frameworks that go beyond providing a measure to inferring a computational representation for a complex system. 
 The former comes with a measure called  \emph{true measure complexity}, the latter with a measure called \emph{statistical complexity}. 
 Since computational mechanics has been developed in  more  detail (see \cite{shalizi_computational_2001}),  we focus on it here. 

 The assumption of computational mechanics is that a complex system is an information-storing and -processing entity. Hence, any structured behaviour it exhibits is the result of a computation. The starting point of the inference method is a representation of the system's behaviour as a string such as, for example, a time sequence of measurements of its location.\footnote{ Of course, measurement is crucial to computational mechanics, and it raises many practical questions left aside here. } The symbols in this measurement sequence generally form a discrete and finite set (for background, see  Appendix~\ref{sec.probtheory}). Once a string of measurement data has been obtained, the regularities are extracted using an algorithm which is briefly explained below, and a computational representation is inferred which reproduces the statistical regularities of the string. Computational representations can, in principle, be anything from the Chomsky hierarchy of computational devices \cite{hopcroft_introduction_2013}, but in concrete examples they usually are finite-state automata. The size of this automaton is the basis for the statistical complexity measure.

The algorithm for inferring the computational representation of a string assumes that a stationary  stochastic process $\{X_t\}_{ t \in {\mathbb Z}}$  has generated the string in question (for a definition of stationary stochastic process, see  Appendix~\ref{sec.probtheory}). 
As a next step, statistically equivalent strings are grouped together. 
Two strings, $\underline{x}$ and $\underline{x}^\prime$,  are statistically equivalent if they have the same conditional probability distribution over the subsequent symbol $a \in \mathcal{X}$:
\begin{align}
\begin{split}
 P(X_{0}=a | X_{t}^{-1} = 
 \underline{x} ) 
= P(X_{0}=a & | X_{t^\prime}^{-1} = 
\underline{x}^\prime
), \\
& \text{ for all}~ a\in\mathcal{X}  ~.
\end{split}
\end{align}
The two strings do not have to be of the same length. The \emph{equivalence class}   of a substring $\underline{x}$ is denoted by  $\epsilon(\underline{x}) $, and  it contains all strings statistically equivalent to string $\underline{x}$, including $\underline{x}$ itself. These classes are  called `causal states', a somewhat unfortunate name since no causality is implied in any strict sense. 
Due to the stationarity of the process, the transition probabilities between the causal states are stationary and form a stochastic matrix. Hence, the  computational representation obtained by this algorithm is a stochastic finite state automaton or, equivalently,  a hidden Markov model \cite{hopcroft_introduction_2013,paz_introduction_1971} and is called $\epsilon$-machine.\footnote{ The inference algorithm is available in various languages, see, for example, \cite{CSSR, python-compmech,kelly2012new-2col}.}

 The  stationary probability distribution $P$ of the $\epsilon$-machine's causal states $s \in \mathcal{S}$, which is the left eigenvector of its stochastic transition matrix with eigenvalue $1$, is used to define the 
 {statistical complexity}, $C_\mu$, of a process: %is the :
\begin{align}
C_\mu := -\sum_{s\in \mathcal{S}} P(s) \log_2 P(s)~,
\end{align}
where $\mathcal{S}$ is the set of causal states. $C_\mu$ is the Shannon entropy of the stationary probability distribution. This reflects the computational viewpoint of the authors since, technically, the Shannon entropy is the minimum number of bits required to encode the set $\mathcal{S}$ with probability distribution $P$. Thus, the statistical complexity is a measure of the minimum amount of memory required to optimally encode the set of behaviours of the complex system. 
It is worth  noting that, for a given string, the statistical complexity is lower bounded by the excess entropy/predictive information (see eq.~\ref{eq.Ipred} above), $C_\mu \geq \textnormal{I}_{\textnormal{pred}}$ (eq.~\ref{eq.Ipred}) \cite{crutchfield_regularities_2003,wiesner2012information,crutchfield2009time}. This mathematical fact agrees with the intuition that a system must store at least as much information as the structure it produces.  
The statistical complexity has been computed for the logistic map  \cite{crutchfield_inferring_1989}, for protein configurations \cite{li_multiscale_2008, kelly2012new-2col}, atmospheric turbulence \cite{palmer_complexity_2000}, and for self-organisation in cellular automata \cite{shalizi_quantifying_2004}.

\cite[p. 24]{crutchfield_calculi_1994} writes that ``an ideal random process has  zero  statistical  complexity.   At  the  other  end  of  the  spectrum,  simple  periodic  processes  have  low  statistical  complexity. Complex processes arise between these extremes and are an amalgam of predictable and stochastic mechanisms.'' This statement, though intuitive, is obscuring the fact that  the statistical complexity  increases monotonically with the order of the string. For a proof consider the following. For a given number of causal states the statistical complexity has a unique maximum at uniform probability distribution over the states. This is achieved by a perfectly periodic sequence with
 period equal to the number of states. When deviations occur, the probability distribution will, in general, not be uniform anymore, and the Shannon entropy and with it the statistical complexity will decrease. On the other hand, increasing the period of the sequence requires an increased number of causal states and, thus, implies a higher statistical complexity. Hence, the statistical complexity scores higher for highly ordered strings than for strings with less order or with random bits inserted. The statistical complexity is a measure of order produced by the system, as well as a measure of memory of the system itself. The strength of the framework of computational mechanics lies in detecting order in the presence of disorder.

 \subsection{Effective Complexity}

{Effective complexity} was introduced by physicists Murray Gell-Mann and Seth Lloyd \cite{gell-mann_information_1996}. 
Gell-Mann and Lloyd's starting point is common to many measures of complexity of that time: the measure should capture the property of a complex system of being neither completely ordered nor completely random. They assume the complex system can be represented as a string of bits; call it $s$. This string is some form of unique description of the system or of its behaviour or the order it produced. The algorithmic complexity (for a definition, see Appendix~\ref{sec.alg.compl}) of this string of bits is a measure of its randomness or lack of compressibility. The more regularities a string has, the lower is its algorithmic complexity. Hence, Gell-Mann and Lloyd consider the algorithmic complexity not of the string itself, but of the ensemble (a term taken from statistical mechanics) of strings with the same regularities as the string in question. Let $E$ be this ensemble of strings with the same regularities. The effective complexity of the string (and thus the system which it represents) is defined as the algorithmic complexity of the ensemble $E$ in which it is embedded as a typical member. (`Typical' is a technical term here, but it captures exactly what we intuitively think `typical' should mean.) Ensemble members $s\in E$ are called typical if $-\log \Pr(s) \approx K_{\mathcal U}(E)$, where  $K_{\mathcal U}(E)$ is the algorithmic complexity of $E$ (see  Appendix~\ref{sec.alg.compl}).\footnote{The idea of replacing entropy (the average of $-\log \Pr(s)$ is an entropy) by algorithmic complexity goes back to Wojciech \cite{zurek_algorithmic_1989}.} Assigning a probability to each string in a set is less arbitrary than it sounds. It has been shown that the probability $\Pr(s)$ of a string $s$ is related to its algorithmic complexity as $- \log \Pr(s) \approx K_{\mathcal U}(s|E)$ where $K_{\mathcal U}(s|E)$ is the algorithmic complexity of the string $s$ given a description of the set $E$. The effective complexity $\varepsilon(s)$ of a string $s$ is defined as the algorithmic complexity of the ensemble $E$ of which it is a typical member, 
\begin{equation}
{\varepsilon}(s) = K_{\mathcal U}(E)~.
\end{equation}
For example, the ensemble of a string which is perfectly random is the set of all strings of the same length. This set allows for a very short description, by giving the length of the strings only. This trick of embedding the string in a set of similar strings exactly achieves what Gell-Mann and Lloyd set out to do. A string with many regularities over many different length scales, which is how we think of a complex system, will be assigned a high effective complexity. Random systems, in their structure or behaviour, will be assigned very low effective complexity. 
According to \cite{gell-mann_information_1996}, the effective complexity can be high only in a region intermediate between total order and complete disorder. However, replacing some of the regular bits in a string by random bits decreases its regularities and hence its effective complexity. Just like the true measure complexity and the statistical complexity, the effective complexity increases monotonically with the amount of order present. 
This places the effective complexity among the measures of order. 
Gell-Mann and Lloyd note that this measure is subjective, since what to count as a regular feature is the observer's decision. The instruction  ``find the ensemble (of a rain forest, for example) and determine its typical members'' leaves too many things unspecified for this measure to be practicable \cite{mcallister_effective_2003}.

\subsection{Logical Depth}

Computer scientist Charles Bennett introduced \emph{logical depth} to measure a system's history \cite{bennett_logical_1995}. Bennett argues that complex objects are those whose most plausible explanations involve long causal processes. This idea goes back to Herbert Simon's influential paper, `The Architecture of Complexity' \cite{simon_architecture_1962}. To develop a mathematical definition of causal histories of complex systems, Bennett replaces the system to be measured by a description of the system, given as a string of bits. This procedure should be very familiar by now. He equates the causal history of the system with the algorithmic complexity of the string (the length of the shortest program which outputs the string; see  Appendix~\ref{sec.alg.compl}). The shorter the program which outputs a system's description, the more plausible it is as its causal history. A program consisting of the string itself and the `print' command has high algorithmic complexity, and it offers no explanation whatsoever. It is equivalent to saying `It just happened' and so is effectively the null-hypothesis. A program with instructions for computing the string from some initial conditions%, on the other hand,
must contain some description of its history and thus is a more explanatory hypothesis.

In addition to considering a program's length as a measure of causal history, Bennett also takes the program's running time into account. A program which runs for a long time before outputting a result signifies that the string has a complicated order that needs unravelling. The definition of logical depth is then as follows. Let $x$ be a finite string, and $K_{\mathcal U}(x)$ its algorithmic complexity. The logical depth of $x$ at significance level $s$ is defined as the least time $T(p)$ required for program $p$ to compute $x$ and then halt where the length of program $p$, $l(p)$, cannot differ from $K_U(x)$ by more than $s$ bits, 
\begin{align}
%\begin{split}
  \text{ D%epth
  }_s(x) := \min_p \{T(p) : l(p) - K_{\mathcal U}(x) \leq s , U(p) = x\}~. 
%\end{split}
\end{align}

Logical depth combines the features of order and history into a single measure. Consider the structure of a protein, for example. One possible program  prints the electron and nuclear densities verbatim, with discretised positional information, which is a very long program running very fast. Another program  computes the structure ab initio by running quantum chemical calculations. This would be a much shorter program but running for a long time. The latter captures the protein's order and history. The real causal history of a protein is, of course, very long, starting with the beginning of life on Earth, or even with the beginning of the universe. The logical depth captures our intuition that complex systems have a long history. A practical problem is that the time point when the  history of a system starts is not well defined. Another aspect which makes it impractical to use is that the algorithmic complexity is uncomputable in principle, although approximations exist such as the Lempel-Ziv algorithm \cite{ziv1977universal}. 

\section{Conclusion}

The study of coupled human and natural systems is vital to our survival. The science that is required to understand and predict phenomena such as climate change and migration is necessarily multidisciplinary. For science to progress developing the right mathematical tools is essential, including in the social sciences. A unifying framework for complexity facilitates this collective endeavour. Furthermore, it facilitates quantitative approaches to issues such as economic growth and income inequality. However, there is confusion about the nature of complexity even within the natural sciences. The analysis in terms of complexity developed in \cite{ladyman2020what} and the above analysis of mathematical measures of these features (which is part of \cite{ladyman2020what}) is intended to bring clarity to the discussion of complexity across all disciplines. The examples cited from economics and neuroscience of purported complexity measures are prominent in the literature. We clarified the interpretation of some of them and incorporated them all into our account as measures of particular features of complexity. Any measure in complexity science can be interpreted as measuring an aspect of one of the features of complexity, and any identifying a feature of complexity that is important for a given system makes it easier to select, modify or design an appropriate measure for it.

%In \cite{ladyman2020what}, we identified the features of a complex system and put them into a framework applicable across the sciences. Here, we reproduced (with small modifications) the chapter on complexity measures. For each feature of a complex system, we gave at least one mathematical means of quantifying it. The exception was the feature `feedback' which is measured indirectly by its effects, and, we believe, usually qualitatively. 

%Any measure that captures complexity in its entirety, if that is even possible, must necessarily an aggregate of measures of some or all the above features of complexity. Whether such an aggregate measure is sensible and useful will depend on the particular system. 

For further and in-depth analysis and discussion we refer the interested reader to our book, ``What is a complex system?'', published with Yale University Press in 2020  \cite{ladyman2020what}.

\section*{Acknowledgement}

We thank Yale University Press for kindly agreeing to us making this chapter publicly available.

\bibliography{core, more}% Produces the bibliography via BibTeX.

%% file: appendix.tex
\appendix

\section*{Appendix} In this Appendix, the mathematical terminology used in the main text is defined, and some more background is given to the mathematical formalism of probability\index{Probability theory} theory, information\index{Information theory} theory, algorithmic\index{Algorithmic complexity} complexity, and network theory. 

\section{Probability Theory}
\label{sec.probtheory}

An \textbf{alphabet} ${\mathcal X}$ is a set of symbols, numeric or symbolic, continuous or discrete, finite or infinite. Symbolic alphabets are discrete; continuous alphabets are numeric and infinite.  $|\mathcal{X}|$ denotes the size of set $\mathcal{X}$. An example of a numeric finite, discrete alphabet is the binary set $\{0, 1\}$; an example of a symbolic, finite alphabet is a set of Roman letters $\{a, b, c, \dots, z\}$. An example of a numeric infinite, discrete alphabet is that of the natural numbers $\mathbb N$, an example of a continuous alphabet is the set of real numbers $\mathbb R$. This book discusses only discrete alphabets. 

A \textbf{discrete random variable} $X$ is a discrete alphabet $\mathcal X$ equipped with a probability\index{Probability distribution} distribution $ P(X) \equiv \{\Pr(X = x),~ x\in {\mathcal X}\}$. We denote the probabilities $\Pr(X = x)$ by $P(x)$ or sometimes, to avoid confusion, by $P_X(x)$. The \textbf{uniform distribution} of a set $\mathcal{X}$ is the distribution $P(x) = 1/|\mathcal{X}|$ for all $x\in \mathcal{X}$. For two discrete random variables, $X$ and $Y$, the joint probabilities $\Pr(X = x, Y = y)$ on  alphabet $\mathcal{X} \times \mathcal{Y}$ are denoted by $P(xy)$ or sometimes, to avoid confusion, by $P_{XY}(xy)$. The joint probability\index{Probability distribution!joint} distribution induces a conditional probability\index{Probability distribution} distribution $P(x|y) \equiv \Pr(X = x | Y = y)$, which is a probability\index{Probability distribution} distribution on $\mathcal X$ conditioned on $Y$ taking particular value $Y = y$. Any joint probability $P(xy)$ can  be written as 
 \begin{align} 
P(xy) = P(x|y)P(y)~.
 \end{align}
 
The \textbf{expectation\index{Expectation value} value} of a discrete numeric random variable $X$, denoted by $\expX{X}$, is defined as 
\begin{align}
  \expX{X}  := \sum_{x\in {\mathcal X}} P(x) x~.
	\end{align} 

Another common notation for the expectation\index{Expectation value} value of $X$ is $\mathbb{E} X$.

The \textbf{variance\index{Variance}} of a numeric random variable $X$ is the average deviation of $X$ from its expectation\index{Expectation value} value. Denoted by $\varX$, it is defined as
	\begin{align}
  \varX :=  \langle (X - \expX{X})^2\rangle~.
	\end{align}

The square root of the variance\index{Variance} of a random variable $X$ is called the \textbf{standard deviation} $\sigma$:
	\begin{align}
	\sigma = \sqrt{\varX}~.%\langle (X - \expX{X})^2\rangle}~.
	\end{align}
    
The ratio of variance\index{Variance} to expectation\index{Expectation value} value is called the \textbf{coefficient of variation}, 	\begin{align}\textnormal{cv} := \frac{\sqrt{\mathrm{Var} X}}{\expX{X}}~.
 	\end{align}
	
The \textbf{covariance\index{Covariance}} of two numeric random variables $X$ and $Y$ is defined as 
\begin{align}
\notag  \text{ Cov}\, XY &:=  \expX{ (X - \expX{X})(Y - \expX{Y})}\\
 \label{eq.cov} 	&= \expX{XY} - \expX{X}\expX{Y}~.
\end{align}

A \textbf{stochastic process} $\{X_t \}_{ t \in T}$ is a sequence of random variables $X_t$, defined on a joint probability space, taking values in a common set $\mathcal X$, indexed by a set $T$ which is often  $\mathbb N$ or $\mathbb Z$ and thought of as time. This book only discusses discrete time processes. A stochastic process is called a \textbf{Markov\index{Markov chain} chain} if $X_t$ (sometimes called `the future') is probabilistically independent of $X_{0}\dots X_{t-2}$ (`the past'), given $X_{n-1}$ (`the present'); in other words, \begin{align}
 P(X_t | X_{0}\dots X_{t-1}) = P(X_t | X_{t-1}) \textnormal{~, for all~}  t\in T~.
\end{align}
 
A stochastic process is \textbf{stationary} if
\begin{align}
\begin{split}
P(X_{t} X_{t+1}\dots X_{t+m}) = &P(X_{t^\prime} X_{t^\prime+1}\dots X_{t^\prime+m}), \\ & \textnormal{ for all}~ t,t^\prime\in T, m\in {\mathbb N}.
\end{split}
\end{align}

A \textbf{hidden Markov\index{Markov model} model}  $\{X_t, Y_t\}_{ t \in T}$ is a stationary stochastic process of two random variables $X_t$ and $Y_t$ which forms a Markov\index{Markov chain} chain in the sense that $Y_t$  depends only on $X_t$, and $X_{t}$  depends only on $Y_{t-1}$ and $X_{t-1}$:
\begin{align}
 P(Y_t | X_{0}\dots X_{t} Y_0 \dots Y_{t-1}) &= P(Y_t | X_{t}) \\ \nonumber &\textnormal{and}\\
 \nonumber P(X_{t} | X_{0}\dots X_{t-1} Y_0 \dots Y_{t-1}) &= P(X_{t} | X_{t-1} Y_{t-1})~,\\
 &\textnormal{~ for all~}  t\in T~.
\end{align}

The graphical representation of a hidden Markov\index{Markov model} model is a directed graph where the states are the outcomes $x\in \mathcal{X}$ of the random variable $X_t$ and the state transitions are labelled by the outcomes $y \in \mathcal{Y}$ of the random variable $Y_t$ and the corresponding conditional probability $P(Y_{t+1} = y, X_{t+1} = x | X_t = x)$.

\section{Shannon Information Theory}\label{sec.info-theory}
%\addcontentsline{toc}{section}{Shannon Information Theory}

In the 1940s, the American engineer Claude Shannon\index{Shannon, Claude}, working for Bell Labs, introduced a mathematical theory of communication\index{Communication}  that is now at the heart of every digital communication\index{Communication} protocol and technology, from mobile phones to e-mail encryption services and wireless networks\index{Networks} \citep{shannon1948mathematical}. Shannon was concerned with defining and measuring the amount of information communicated by a message transmitted over a noisy telegraph line. He saw a message as communicating  information if the receiver of the message could not predict with certainty which message out of a set of possible ones she would receive. By setting the amount of information communicated by a message $x$ as proportional to its inverse log probability  $1/\log P(x)$, Shannon axiomatically derived a measure of information, now called Shannon entropy. 
The \textbf{Shannon entropy}, a function of a probability\index{Probability distribution} distribution $P$ but often written as a function of a random variable $X$, is defined as follows \citep{cover_elements_2006}:
\begin{align}
 H(X) := - \sum_{x\in {\mathcal X}} P(x) \log P(x)~,
\end{align}
where the $\log$ is usually base $2$ and $0 \log 0 := 0$. 
The equivalent definition 
\begin{align}
H(P) :=  - \sum_{i=1}^n p_i \log p_i~,
\end{align}
where $P = \{p_1, p_2, \dots, p_n\}$, 
makes it explicit that $H$ is a function of the probabilities alone, independent of the alphabet $\mathcal{X}$. This book  discusses only the entropy\index{Entropy} of finite probability\index{Probability distribution} distributions, but the definition of the Shannon entropy\index{Entropy!Shannon entropy} extends to infinite but discrete, as well as to continuous probability\index{Probability distribution} distributions. Taking the logarithm to base $2$ is a convention dating back to Shannon, due to a \emph{bit} being the essential unit of computation\index{Computation}. For a given set of messages $\mathcal{X}$, the Shannon entropy\index{Entropy!Shannon entropy} is maximum for the uniform distribution and proportional to the logarithm of the total number of messages. This illustrates that the Shannon entropy\index{Entropy!Shannon entropy} is a measure of randomness\index{Randomness}. If one of the messages has probability 1 and the others have probability 0, then the message is perfectly predictable, and the Shannon entropy\index{Entropy!Shannon entropy} is zero. The Shannon entropy\index{Entropy!Shannon entropy} is also precisely the expectation\index{Expectation value} value of the function  $ 1 / \log P(x)$. 

The \textbf{joint entropy} of $n$ random variables $X_1, \dots, X_n$ with joint probability\index{Probability distribution!joint} distribution $P(X_1X_2\dots X_n)$ is defined as 
\begin{align}
\begin{split}
  H(X_1X_2\dots X_n) := -\sum_{\substack{x_1\dots x_n  \in \\ \mathcal{X}_1\times \dots \times \mathcal{X}_n}} & P(x_1 x_2 \dots x_n) \\
  & \cdot \log P(x_1 x_2 \dots x_n)~.
\end{split}
\end{align}

Consider two random variables $X$ and $Y$ and joint probability\index{Probability distribution!joint} distribution $P_{XY}$. The \textbf{ conditional  entropy} of $X$ given $Y$, $H(X | Y)$, is defined as 
\begin{align}
  H(X|Y) &:= -   \sum_{xy \in \mathcal{X\times Y} } P_{XY}(xy) \log P_{XY}(x|y)~.
\end{align}

The \textbf{ entropy\index{Entropy!entropy rate} rate} of a  stochastic process $\{X_t\}_{ t \in T}$  is defined as 
\begin{align}
h &= \lim_{n\to \infty}\frac{1}{n} H(X_{1}X_{2} \dots X_n  )  ~.
\end{align}
A different definition of entropy\index{Entropy!entropy rate} rate is as follows:
\begin{align}
h^\prime &= \lim_{n\to \infty} H(X_n | X_{1} \dots X_{n-1}) ~.
\end{align}
For stationary stochastic processes, $h = h^\prime$. 
The entropy\index{Entropy!entropy rate} rate $H(X_n | X_{1} \dots X_{n-1}) $, for finite $n$, is denoted by $h_n$.

Shannon introduced the \textbf{mutual information\index{Mutual information}} as a measure of correlation\index{Measures!of correlation} between two random variables $X$ and $Y$, defined as follows:

 \begin{align}\label{eq:I}
  I(X;Y) :=  \sum_{xy\in \mathcal{ X\times Y}} P_{XY}(xy) \log \frac{ P_{XY}(xy)}{P_X(x)P_Y(y)}~.
 \end{align}

 The mutual information\index{Mutual information} is a measure of the predictability\index{Measures!of predictability} of one random variable when the outcome of the other is known. Note that the mutual information\index{Mutual information} is symmetric in its arguments and hence measures the amount of information `shared' by the two variables. The mutual information\index{Mutual information} is a general correlation\index{Correlation function} function for two random variables, measuring both linear and non-linear correlation\index{Correlation}s. In contrast to the covariance\index{Covariance} and many other correlation measure\index{Measures!of correlation}s, it is also applicable to non-numeric random variables such as the distribution of words in an English text\index{Distribution!of words in English} or the distribution of amino acids in a DNA sequence. This is one reason why it is widely used in complex systems research.

\section{Algorithmic\index{Algorithmic Information Theory} Information Theory}
\label{sec.alg.compl}
%\addcontentsline{toc}{section}{Algorithmic Information Theory}

A mathematical formalisation of randomness\index{Randomness} and information without reference to probabilities was developed independently by the Soviet mathematician Andrey Kolmogorov and the American mathematicians Ray J. Solomonoff and Gregory Chaitin in the 1960s. They considered information as a property of a single message, rather than of a set of messages and their probabilities. A message is a string of letters from an alphabet, such as the Roman alphabet or the binary characters 0 and 1. An example of a string is `{Hello, World!}'. 
The string is composed of letters from the Roman alphabet and from a set containing the comma and space characters and the exclamation mark. 

The \textbf{algorithmic information} content of a string is, roughly speaking, the length of the shortest computer program which outputs the string. For the string `{Hello, World!}', this is probably a program of the form `{print(``Hello, World!'')}' which has roughly the same length as the string itself. However, for a string of 10,000 zeros and ones alternating, the shortest program is shorter than the string itself, and the string is called `compressible'. The notion of compressibility is meaningful only with longer strings. Only perfectly random strings are completely incompressible, therefore algorithmic information is a measure of randomness\index{Measures!of randomness}. It can be confusing that the term `information' is used for randomness\index{Randomness}, but one may think of randomness\index{Randomness} as the amount of information which has to be communicated to reproduce the string `exactly', irrespective of how interesting the string is in other respects. 
 
The precise definition of {algorithmic  information} is as follows \citep{li_introduction_2009}. Consider a string $x$, a computing device $\mathcal U$ and programs $p$ of length $l(p)$. The algorithmic information of the string, $K(x)$, is the  length of the shortest program $p$, which, when fed into a machine $\mathcal U$, produces output  $x$, ${\mathcal U}(p) = x$, 
\begin{align}
 K_{\mathcal U}(x) = \min_{p:{\mathcal U}(p) = x} l(p)~.
\end{align}

The minimisation is done over all possible programs. There is a fundamental problem with carrying out the minimisation procedure: whether an arbitrary program will finish or run forever cannot be known in general. This is called the halting problem. It is one of the deepest results in computer science, and due to the British mathematician Alan Turing\index{Turing, Alan}. As a consequence, the algorithmic information is not computable in principle, though it can often be approximated in practice. Other names for algorithmic information are `algorithmic\index{Algorithmic complexity} complexity' or `Kolmogorov complexity'.

The fundamental insight of  Kolmogorov, Solomonoff and Chaitin is that the minimum length of a program is independent of the computing device on which it is run (up to some constant which is independent of the string).  Hence, the definition of algorithmic information refers to a universal computer, which is a fundamental notion introduced by  Alan Turing\index{Turing, Alan} in the 1940s. Algorithmic information is therefore a `universal' notion of randomness\index{Randomness} for strings because it is context- (machine-) independent. On the other hand, the Shannon entropy\index{Entropy!Shannon entropy} is context-dependent, since it may assign different amounts of information to the same string when it is embedded in different sets with different probabilities.

The length is not the only important parameter of a program; its running time is of equal importance. There are very short programs that take a long time to run, while the print program might be long but finished very quickly. This trade-off is relevant to the measures of complexity, which include the well-known logical depth\index{Logical depth}.\\

\section{Complex Networks\index{Networks}}
%\addcontentsline{toc}{section}{Complex Networks}
%\begin{framed}
%\textcolor{blue}{image of network?}
%\end{framed}

A \textbf{network}, or a graph, is a set of nodes and, for simplicity here, there is at most one edge between any ordered pair of nodes. Nodes and edges are also called vertices and links, respectively. In a \textbf{directed network} each edge has a directionality, beginning at one node and ending at another. In an undirected network there is no such distinction between the start and end node of an edge. An example of an undirected network is the Internet. The servers are nodes, and edges between them are the physical wirings. An example of a directed network is an ecological food\index{Food webs} web. Two animals are linked if one of them feeds on the other so that a predator has a directed edge to its prey.

The \textbf{degree} of a node in a network is the number of edges attached to it. In a directed network, one distinguishes between in-degree and out-degree. The in-degree of a node is the number of edges directed to the node, and the out-degree is the number of nodes directed away from it.

Mathematically, a network of $n$ nodes is represented by an \textbf{adjacency matrix\index{Matrix!adjacency}}, $A$, which is an $n \times n$ matrix\index{Matrix!adjacency} where each non-zero entry $A_{ij}$ represents an edge from node $i$ to node $j$ \citep{newman_networks:_2010}. In an unweighted network the $A_{ij}$ are $ 1$ if an edge exists from node $i$ to node $j$ and $0$ otherwise. A \textbf{weighted network} assigns a real number to each edge, $A_{ij} \in {\mathbb R}$. Such weights could, for example, represent the volume of data traffic between two servers. In an undirected network, $A_{ij} = A_{ji}$ since $A_{ij} $ and $ A_{ji}$ refer to the same object. 
        The in-degree of a node $i$ is the number of non-zero entries in the $j^{th}$ column of $A$. The out-degree of a node $i$ is the number of non-zero entries in the $i^{th}$ row of $A$. In an undirected network, these numbers are equal. 
  
The \textbf{degree\index{Distribution!degree distribution} distribution} of a network is the frequency\index{Distribution!frequency distribution} distribution over node degrees. A uniform degree\index{Distribution!degree distribution} distribution, for example, means that nodes of degree $1$ are equally likely as nodes of degree $n$. A \textbf{path} is a sequence of nodes such that every two consecutive nodes in the sequence are connected by an edge. In a directed network the nodes have to be connected by edges that all point in the forward direction. The \textbf{ path length} is the number of edges traversed along the sequence of a path. The \textbf{shortest path} between two nodes is the sequence with the minimum number of traversed edges to get from one node to the other. The average shortest path is the sum of all shortest path lengths divided by their number. The \textbf{diameter} of a network is the longest of all shortest paths.